\documentclass[10pt,twocolumn,letterpaper]{article}

\usepackage{cvpr}
\usepackage{times}
\usepackage{epsfig}
\usepackage{graphicx}
\usepackage{amsmath}
\usepackage{amssymb}
\usepackage{algorithm}
\usepackage{algorithmic}
\usepackage[pagebackref=true,breaklinks=true,letterpaper=true,colorlinks,bookmarks=false]{hyperref}

\cvprfinalcopy 

\def\cvprPaperID{****} 

\ifcvprfinal\fi
\begin{document}

\title{Understanding the efficacy, reliability and resiliency of computer vision techniques for malware detection and future research directions}

\author{Li Chen\\
Data/Research Scientist, Security and Privacy Lab, Intel Labs\\
}

\maketitle
\begin{abstract}
  My research lies in the intersection of security and machine learning. This overview summarizes one component of my research: combining computer vision with malware exploit detection for enhanced security solutions. I will present the perspectives of efficacy, reliability and resiliency to formulate threat detection as computer vision problems and develop state-of-the-art image-based malware classification. Representing malware binary as images provides a direct visualization of data samples, reduces the efforts for feature extraction, and consumes the whole binary for holistic structural analysis. Employing transfer learning of deep neural networks effective for large scale image classification to malware classification demonstrates superior classification efficacy compared with classical machine learning algorithms. To enhance reliability of these vision-based malware detectors, interpretation frameworks can be constructed on the malware visual representations and useful for extracting faithful explanation, so that security practitioners have confidence in the model before deployment. In cyber-security applications, we should always assume that a malware writer constantly modifies code to bypass detection. Addressing the resiliency of the malware detectors is equivalently important as efficacy and reliability. Via understanding the attack surfaces of machine learning models used for malware detection, we can greatly improve the robustness of the algorithms to combat malware adversaries in the wild. Finally I will discuss future research directions worth pursuing in this research community.
  
  
  
\end{abstract}

My research explores and addresses the current gaps within the intersection of security and artificial intelligence. In this paper I will provide an overview on extending the advances in computer vision research to better effectively detect malware. 

As the diversity and volume of malware continue to increase rapidly, intelligent algorithms are in high demand to effectively detect malware. Two main approaches for malware detection are static code analysis and dynamic code analysis. Static code analysis dissembles the code and discovers malicious patterns in the control flow executable without executing the applications or monitoring their run time behaviors. Although it is fast, static analysis is not effective against obfuscated code or morphed malware whose malicious signature may not exist in the known malware signature database.  Dynamic code analysis executes the code to reveal run-time behaviors and persists more effectiveness against packed or encrypted code, but it can consume more time and still miss detection if triggering conditions are not met. 

What follows is heavy emphasis on excellent feature construction including binning the type of function calls, counting the number of loops, and applying language processing techniques on either the disassembled code or the run-time sequence calls. The extracted features can be very high-dimensional, so dimension reduction methods such as principal component analysis or multi-dimensional scaling are needed due to the curse of dimensionality. However drawbacks remain, as feature extraction does not comprehend the overall structural information of the malware and selecting the right dimension for dimension reduction is also proven to be difficult.

Computer vision, on the other hand, provides a unique perspective for malware detection without dissembling the code, executing the code or extracting the features. A malware binary can be directly converted to pixel values between 0 and 255 \cite{nataraj2011malware, makandar2015malware, yue2017imbalanced, chen2018henet, chen2018deep}. By visual inspection of application binaries plotted as grey-scale images: there exists textural and structural similarities among malware from the same family and dissimilarities between malware and benignware as well as across different malware families. 

How we can adequately utilize the vision-driven information encourages further exploration and investigation of merging advances in computer with malware exploit detection. Here I will explain my research in this domain via three perspectives, namely \textit{(i)} efficacy, \textit{(ii)} reliability and \textit{(iii)} resiliency. 

\section{Efficacy}

The vision-based approach for malware classification consumes the structural and textural information of malware or benign applications as a whole, when the binaries are directly converted to pixel values between 0 and 255. Such an approach provides a visualization on the abstract malware samples. 

In \cite{chen2018deep}, I propose deep transfer learning for static malware classification, where I augment the grey-scale malware images into RGB-channels, and apply transfer learning on the malware dataset. The pre-trained deep neural networks such as Inception, VGG or ResNet are obtained from natural images from ImageNet database, so that the models contain significant amount of learned features from large quantities of images. Transfer learning from the natural image domain to the malware image target domain minimizes the efforts to search for the optimal neural network architecture or the best parameter sets, accelerates the training time on the malware dataset, while still maintaining high classification accuracy and false positive rate. 

In all the real data experiments in \cite{chen2018deep}, the proposed method outperforms with the highest classification accuracy, lowest false positive rate, highest true positive rate and highest $F_1$ score compared with all other selected classical machine learning algorithms such as shallow fully connected neural networks (shallow NN), naive Bayes, 5-nearest neighbor (5NN), linear discriminant analysis (LDA), random forest, XGB, support vector machine with linear kernel (SVM-linear), support vector machine with radial kernel (SVM-radial), and also outperforms training-from-scratch scheme. A performance table is presented in Table \ref{tab:dataset2}. 

Furthermore we extend deep transfer learning for dynamic exploit detection \cite{chen2018henet}, where we convert the control flow packets generated from Intel Processor Trace into time series of images, propose a hierarchical ensemble neural network (HeNet) via deep transfer learning for dynamic return-oriented-programming attacks and show its highest classification accuracy with lowest false positive rate compared with commonly used machine learning algorithms such as random forest, nearest neighbor, naive Bayes. Indeed, vision-based transfer learning techniques on malware images not only save tremendous efforts for manual feature engineering, but also possess superior performance for malware classification tasks.

\begin{table}[t]
\begin{center}
\begin{tabular}{ |l|c|c| c| } 
 \hline
Algorithm & Accuracy & $\overline{FPR}$ & $\overline{TPR}$ \\ \hline 
Proposed method & \textbf{98.13\%} & \textbf{0.237\% } & \textbf{96.63\%} \\ 

TFS via shallow NN & 82.41\% & 2.551\% & 59.38\%\\ 

Naive Bayes  & 74.23\%& 3.116\% &73.06\%\\ 
5-nearest neighbor  & 95.31\%&  0.602\% & 85.75\%\\ 
LDA $\circ$ PCA & 76.45\% & 3.023\%  & 63.86\%\\ 
Random forest$\circ$ PCA  & 95.73\%& 0.548\% &  84.26\%\\ 
XGB $\circ$ PCA & 96.01\%& 0.514\%  &  85.80\%\\ 

SVM-linear$\circ$ PCA  & 86.35\% & 1.799\%  & 72.71\%\\ 
SVM-radial$\circ$ PCA  & 86.26\% & 1.975\% &  72.14\%\\ 


\hline
\end{tabular}
\end{center}
\caption{Comparison of algorithm performance on Microsoft Malware Dataset 2015\cite{ronen2018microsoft}. The proposed method in \cite{chen2018deep} achieves the highest classification accuracy, highest average true positive rate and lowest false positive rate. }
\label{tab:dataset2}
\end{table}

\section{Reliability}

Despite the effectiveness of the computer vision based methods for malware classification, understanding the reason why the image-based transfer learning methods makes such predictions on the malware images is critical for security researchers and practitioners. The interpretations will generate valuable insights to triage malware families and enhance the practitioners' trust to the model. Hence an effective model for deployment need not only the best classification performance but also the best reliability from being able to explain its predictions.

\begin{figure}[b]
\centering
\includegraphics[width=0.5\textwidth]{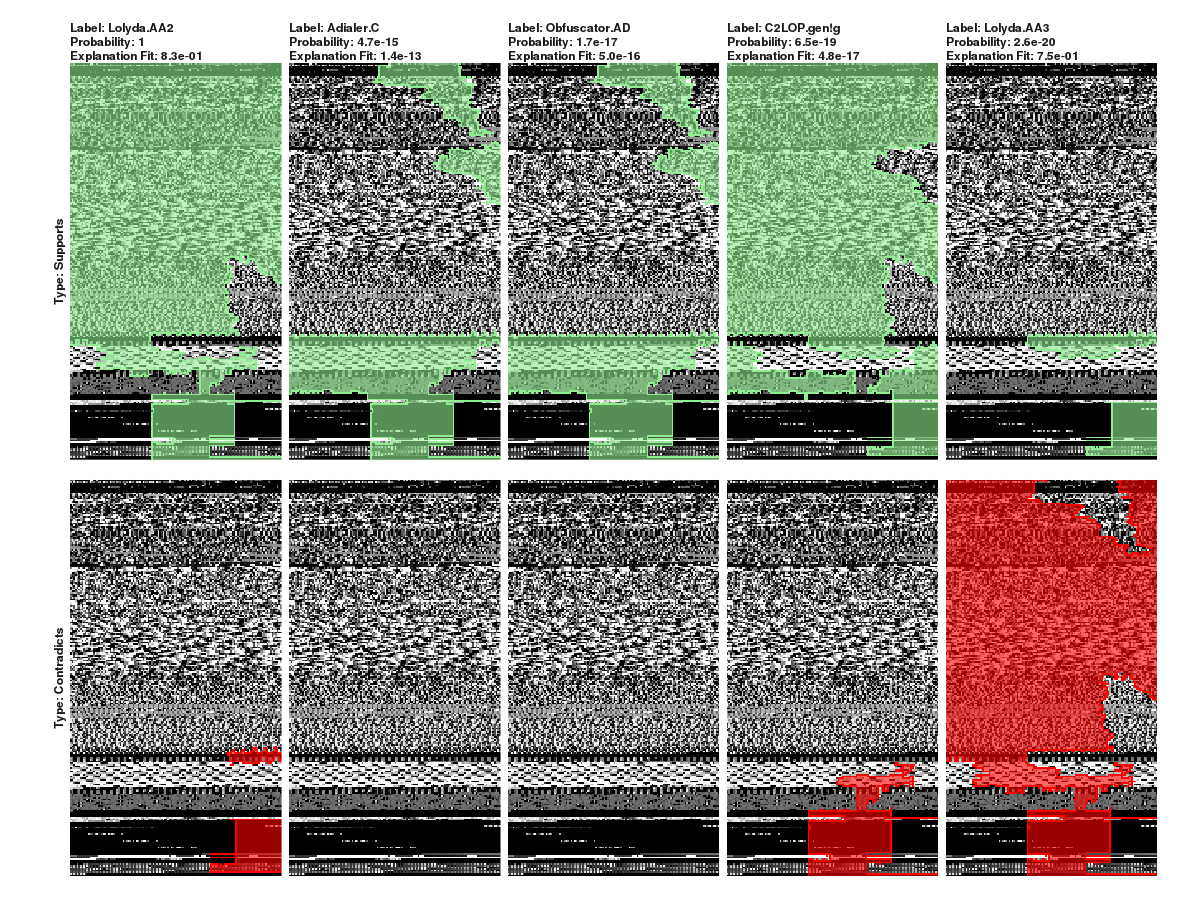}
\caption{\label{fig:interpretation_sample} Visual interpretation of what image-based malware classifier sees. The red regions indicate the pixel regions which the model does not believe they contribute to the prediction. Most area are plotted as green. On the other hand, the top 5-th prediction is Lolyda.AA3, and most of the regions are red, indicating the model sees the least of Lolyda.AA3 family in this malware image.}
\end{figure}

In \cite{chen2018deep}, I propose to extend the local-interpretable model-agnostic explanation approach \cite{ribeiro2016should} to identify which regions in the malware binary contribute to prediction by the neural networks. An example is seen in Figure \ref{fig:interpretation_sample}. Such interpretability highlights the advantage of approaching the malware problem from computer vision direction, so that interpretation becomes concrete as to indicate the actual locations of potential malicious signals. Security practitioners, based on the algorithmic interpretation finding, can check the code and verify whether the ML-identified locations contain the malicious signatures unique to certain families. This direction provides one step closer to uncover reasoning behind black-box deep learning algorithms for malware detection.

\section{Resiliency}
There are always adversaries who intentionally want to bypass malware detection. The importance of studying the attack surface of machine learning algorithms for malware detection helps improve the security and resiliency of the malware detection systems. 

In a case study \cite{chen2018towards},we examine the robustness and resiliency of machine learning based ransomware detection systems. Specifically we propose to synthesize dynamic ransomware behaviors via the auxiliary generative adversarial network (AC-GAN) and demonstrate that the generated malicious behaviors can greatly reduce the efficacy of black-box ransomware classifiers.

GANs are primarily used in computer vision to generate natural images that seem real to the human eyes and their training process can be terminated when the generated images resemble the real ones. However, the inputs in our case study are dynamic ransomware execution logs, so we modify the training termination criterion based on the loss function of the discriminator. To avoid mode collapsing issues in training, we segment the trace logs and employ transfer learning from GANs applied on natural images to enable faster convergence and better quality sample generation. We further propose a set of adversarial quality metrics to quantify the generalized maliciousness in the generated dataset. Our discoveries indicate a broad attack surface on even black-box ML-based malware detectors and advocates adversarial training to enhance the robustness of the system. The case study emphasizes another critical vector for security-based machine learning usage: how to establish model resiliency to defend against carefully crafted adversarial malware attacks. 


\begin{table}[t]
\centering
\begin{tabular}{ | l | c | r |c|}
\hline
  Classifier & Accuracy & FPR & Adv. Dec \\ \hline
  Text-CNN & 0.9890 & 0.0300&0.0000\\ 
 
 XGB $\circ$  Text-CNN & 0.9841 & 0.0032 &0.1273\\
 LDA  $\circ$ Text-CNN   &  0.9865 &0.0494&0.0000\\
 
 SVM-linear  $\circ$ Text-CNN   &0.9881  & 0.0432&0.0000\\
 
  SVM-radial $\circ$  Text-CNN   & 0.9897  & 0.0228 &1.0000\\ \hline
\end{tabular}
\caption[Performance table 1]{Classification performance on the test set. Text-CNN achieves the best classification performance on the raw ransomware behavior dataset. After composing with Text-CNN, all other classifiers' performance significantly improve. Detection results on the generated malicious samples show four of the five highly effective classifiers degrade severely in performance and only one classifier maintains resiliency against attacks. This quantifies the attack surface for these ML-based ransomware detection algorithms.  }
\label{tab:ml_performance}
\end{table}

\section{Future Research Directions}
The recent advances of computer vision motivate novel cybersecurity measures. Below are a few research directions worth considering within this research community.

\begin{itemize}
    \item \textbf{Semi-supervised learning} 
    Semi-supervised algorithms are greatly desired to fit the practical challenges of data without ground truths or evolving malware families. We previously proposed model-based semi-supervised learning for dynamic Android malware detection \cite{chen2017poster}. Extending the model-based approach to image-based malware samples can be valuable to address the issues mentioned above.
    \item \textbf{Interpretability} We will continue the study of interpretability and explainability of deep learning models for image-based malware detection. 
    We plan to investigate the schemes of establishing an overall trustworthy score for the deep learning model and use such a score for model selection for deployment in cyber-security applications.
\end{itemize}

{\small
\bibliographystyle{ieee}
\bibliography{egbib}
}

\end{document}